# Empowering Manufacturers with Privacy-Preserving AI Tools: A Case Study in Privacy-Preserving Machine Learning to Solve Real-World Problems

Xiaoyu Ji (Purdue University), Jessica Shorland (Harvard University), Joshua Shank (Harvard University), Pascal Delpe-Brice (Harvard University), Latanya Sweeney (Harvard University), Jan Allebach (Purdue University), and Ali Shakouri (Purdue University)

**Abstract:** Small- and medium-sized manufacturers need innovative data tools but, because of competition and privacy concerns, often do not want to share their proprietary data with researchers who might be interested in helping. This paper introduces a privacy-preserving platform by which (1) manufacturers may safely share their data with researchers through secure methods, (2) those researchers then create innovative tools to solve the manufacturers' real-world problems, and then (3) provide tools that execute solutions back onto the platform for others to use with privacy and confidentiality guarantees. We illustrate this problem through a particular use case which addresses an important problem in the large-scale manufacturing of food crystals, which is that quality control relies on image analysis tools. Previous to our research, food crystals in the images were manually counted–which required substantial and time-consuming human efforts—but we have developed and deployed a crystal analysis tool which makes this process both more rapid and accurate. The tool enables automatic characterization of the crystal size distribution and numbers from microscope images while the natural imperfections from the sample preparation are automatically removed; a machine learning model to count high resolution translucent crystals and agglomeration of crystals was also developed to aid in these efforts. The resulting algorithm was then packaged for real-world use on the factory floor via a web-based app secured through the originating privacy-preserving platform, allowing manufacturers to use it while keeping their proprietary data secure. After demonstrating this full process, future directions are also explored.

# 1. INTRODUCTION

Artificial intelligence (AI) has revolutionized many industries in the last decade. Instead of extensive analytical, numerical, or statistical models, one can use training sets to extract trends and optimize agile operations. However, so far the power of AI has been used to a limited extent for manufacturing [30]. The major challenge in the use of AI for manufacturing is that current AI techniques work well with large and well-studied training sets which are only available in large-scale manufacturers for a narrow range of products and for a few dominant players. In manufacturing, however, there is a diversity of products, tools, and applications. Manufacturing is highly cost competitive. Aside from an example such as integrated circuits, it is not possible to have a perfect and tightly controlled manufacturing environment. Processing variability and input material variability are inherent. In a typical factory, one can see legacy equipment next to state-of-the-art tools. Also, each product is slightly different.

In addition, although small and medium-sized enterprises (SMEs) comprise 98% of US manufacturing and ~45% of employment, they lack the dedicated data analytics workforce to benefit from the latest advances in AI. Aggregation of data from small manufacturing runs or different SMEs requires a more distributed AI that is co-developed together with privacy and confidentiality policies. Privacy and confidentiality policies are critical as the need for proprietary secrecy is a deeply ingrained part of manufacturing competitiveness. These privacy concerns have been a major barrier for companies who make goods to fully benefit from digital commons not only between competitors but also for different players in a vertical supply chain.

In this paper, we take (1) a real-world obstacle for a manufacturing process (in this case, crystal counting), and (2) show how the data produced by that process might be shared with researchers (in this case, scientists at Purdue University) to come up with an innovative solution. However, most importantly, we do so by (3) using a cutting-edge data sharing environment which guarantees that the manufacturer's proprietary means of production are not compromised, therefore demonstrating a method which might level the playing field for SMEs more broadly and in other use cases in terms of data sharing and the eventual benefits of AI-based solutions on large sets of training data.

*1.1 Image analytics and machine learning in manufacturing*

In order to understand the innovations made by the use of a secure data commons, we must first examine the use case of image analytics in the control of solution crystallization process in the food and pharmaceutical industries. Image analytics and machine learning are essential in manufacturing quality control, particularly detection and process monitoring [4-7]. These methods enable efficient and accurate detection of flaws in product samples, improving quality control. In the food and pharmaceutical industry, hyperspectral imaging has emerged as a valuable tool for assessing product quality and safety [1]. This imaging process has also been applied to monitor and control the quality of pharmaceutical tablets through chemical imaging and spectroscopy techniques [2]. This is driven by that industry's shift towards quality-by-control (QbC) and process analytical technology (PAT) to ensure consistent product quality [3].

We have witnessed significant progress in the control of solution crystallization processes in the food and pharmaceutical industries, driven by the need for improved and consistent quality of the products [23]. Real-time monitoring of crystal size distributions (CSD) can provide valuable insights into the solidification of pharmaceutical ingredients and the conditions under which this process occurs [24]. Moreover, imaging techniques have been applied to study stirred tank crystallizers to analyze images and obtain crystal size and shape data [20]. An automated measuring methodology for crystal size in sweetened condensed milk using ImageJ software has also been proposed [11]. In addition, in-situ crystal size and shape identification using image analysis has been proposed for monitoring crystallization processes, highlighting the importance of real-time imaging systems for crystal size measurement and shape identification [22]. More recently, deep learning-based image segmentation and classification have been utilized for in-line measurement of multidimensional size, shape, and polymorphic transformation of pharmaceutical crystals [16]. The imaging techniques have enabled real-time monitoring and control of critical crystal properties. However, this requires off-line calibration based on the data distribution [12].

In summary, combining machine learning and image analytics in manufacturing quality control has shown promising results in defect detection, process monitoring, and quality assurance. These advanced techniques have the potential to significantly improve efficiency, accuracy, and overall product quality that are desirable in profit-driven manufacturing processes. In the following, we introduce the background of image analytics on microscopic images during food processing in Section 2. Section 3 describes the methods including the development and implementation of the image analytics program, and the new privacy-preserving image analysis tool. Section 4 shows the results of our proposed tool and how it may be employed in real-world practice. We will then discuss the feedback and further opportunities of our tool in section 5.

## 2. BACKGROUND

Seed count is an early stage of the food crystal manufacturing pipeline where crystal seeds are sparsely distributed and, during this stage, crystal distribution is an important measurement for product quality control. If a specific standard is not met, halting the process at the early stage saves the time, effort, and expense of initiating the subsequent manufacturing stages. To count the number of crystals in a liquid sample, microscopic image analytics is an efficient method. Previous works related to food crystal image analytics such as crystals from milk, cheese, and sugar [11, 28, 29] can be separated into traditional image processing methods and machine learning methods. Early works utilized developed image processing software such as ImageJ to count crystals [11]. The limitations of similar ready-to-use software are fixed parameters algorithms and lack of latest image processing techniques. As the libraries of programming languages become publicly available, image processing pipelines can be more adaptive to the specific data distribution [15]. However, machine learning models, which have made remarkable achievements in recent years [16, 17, 18, 19], have shown better performance than the traditional image processing algorithms due to their self-learning ability.

## 3. METHODS

The following section introduces the crystal counting algorithm at the core of this research; a significant advancement in automating the analysis of food crystal images which also represents a breakthrough in accuracy and efficiency for quality control processes in manufacturing. The methodology outlined will describe the steps involved in developing and refining the algorithm, including the imaging techniques used to capture data, the image processing pipeline designed for different camera resolutions, and the integration of machine learning models for crystal detection and analysis. Additionally, the process of deploying this tool in a privacy-preserving environment using MyDataCan will be detailed, demonstrating how the algorithm is applied in real-world manufacturing settings in a way that keeps the manufacturer's proprietary data fully secure.

*3.1 Dataset*

In this subsection, we introduce the imaging method used to capture the dataset under different settings. The stage of the seed count is the first stage of the manufacturing pipeline, and our image analysis algorithm focuses on this stage and automatically generates analysis results to save the human efforts of manual counting and yield more consistent results independent of operator-to-operator variability. The samples of the product are in the form of liquid and are dropped on one glass slide and covered with another, and a microscope captures the image with a fixed focal length. Because there is extra air and dust between the two glasses, we observe air bubbles and air dust noises in the images. During the past year, the manufacturer updated the microscopy device once to achieve a higher resolution. Figure 1 shows two examples of old and new camera images. The old camera image has a resolution of 1280 pixels by 1024 pixels, while the new camera image is 2048 pixels by 1536 pixels.

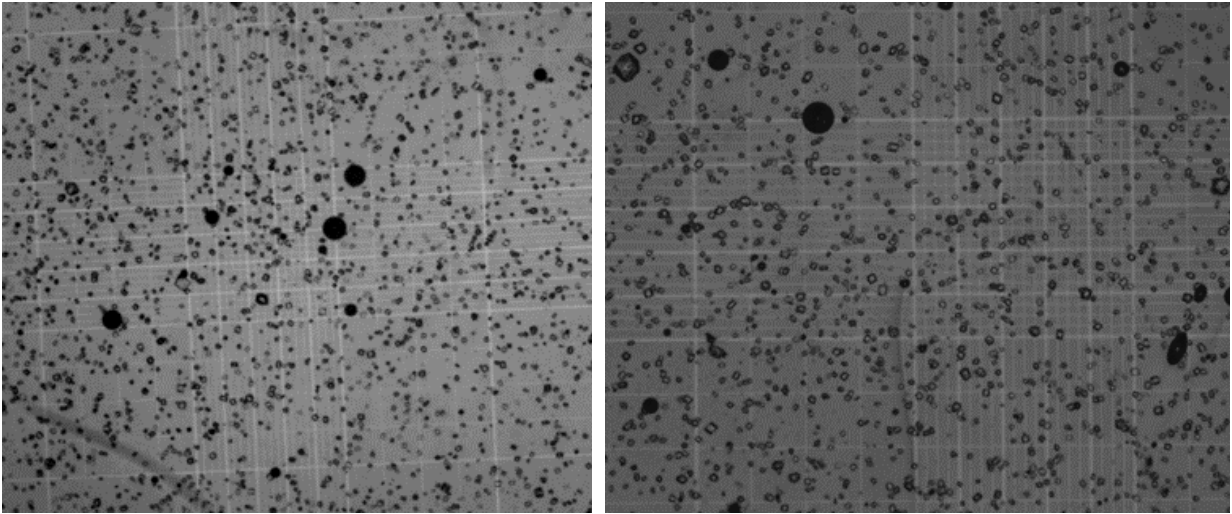

Figures 1 and 2: Food crystal image examples: old camera image (left), new camera image (right). These were not obtained from the same seed count crystal samples.

### 3.2  Image analysis algorithms and models

We developed two different methods to analyze the images from old and new cameras because the resolution difference affects the performance of the program. We first briefly introduce the image processing pipeline based on a baseline method [15] that works with the old camera data. Compared to the related works using ImageJ, an image processing tool to analyze crystal images [1, 12], our pipeline is more adaptive to our own data.

The pipeline can be separated into five parts, as shown in Figure 3. There are two modules that are different from the previously proposed baseline method [15]: the image preprocessing module (whose modifications are shown in Figure 4), and the small cluster removal module (whose further addition is shown in Figure 5).

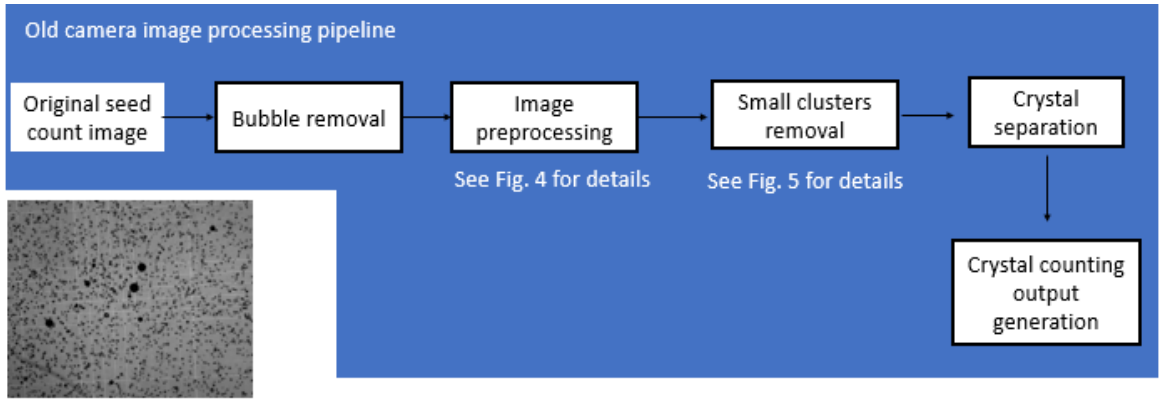

Figure 3: Old camera image processing pipeline.

The image preprocessing module is designed to binarize the image, retaining only the boundaries of the food crystals. Due to uneven lighting under the microscope camera, using a global threshold for binarization can cause edges to be lost in darker areas. To address this, adaptive thresholding [14] is applied, estimating the local threshold based on the surrounding pixels (Figure 4, bottom row). The top row of Figure 4 shows the process of generating a crystal mask, which uses edge detection and additional morphological operations like adaptive smoothing and closing. These operations adjust their kernel size to account for different lighting conditions in the original image. The adaptive smoothing removes background grid noise that may interfere with the Canny edge detection algorithm, while the adaptive closing prevents background noise from forming boundaries, reducing noise after the flood operation.

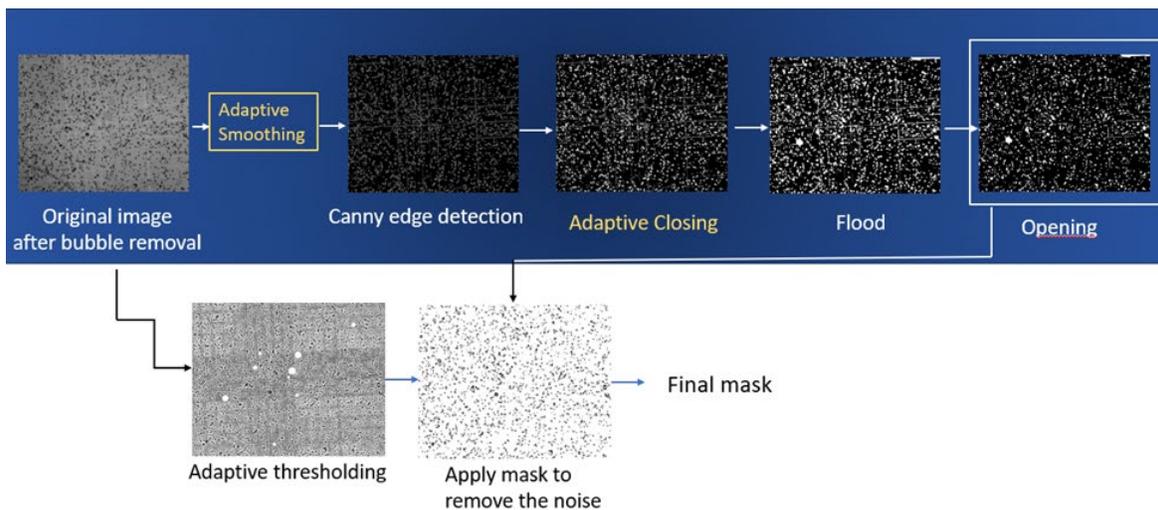

Figure 4: Image preprocessing pipeline.

The small cluster removal module is added to remove noise detected from the image preprocessing output and follows the assumption that each crystal has an opening inside. Experimental thresholds for the small clusters were set but the internal opening was hardly observed. Figure 5 illustrates the three conditions of crystal clusters: no opening, one opening and multiple openings. Convex full detection is added to the cluster with no openings to avoid crystals with one or more faint edges. The threshold for the first two conditions is 20 pixels because such clusters contain at most one crystal. If there are multiple openings in a cluster, we assume there is an overlap between crystals, and the minimum size is set as 5 pixels.

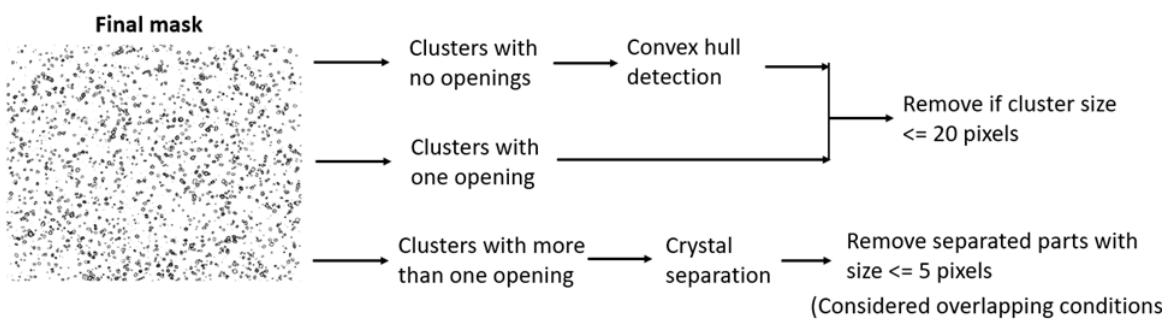

Figure 5: Small cluster removal.

Due to the higher resolution of the new camera, the detailed grayscale color of each facet of translucent crystals is observed in the image, and the background grids are more evident than old camera images. It is difficult to use established methods of image processing due to the fixed nature of the manufacturing pipeline. It is challenging to tackle this with the previous image processing method because of the fixed parameters. StarDist [19] is our baseline architecture, which performed better than other baselines (Figure 6). A possible solution was hypothesized in which the cell segmentation problem could be solved by replacing the rectangle bounding box with a polygon shape. As shown in Figure 8, the backbone is the Unet architecture [26], and there are two convolutional feature layers after the backbone network. The 32-dimensional distance between the boundary and the center of the polygon is the probability of an instance. Non-Maximum Suppression (NMS) [27] eliminates overlapping redundant predictions.

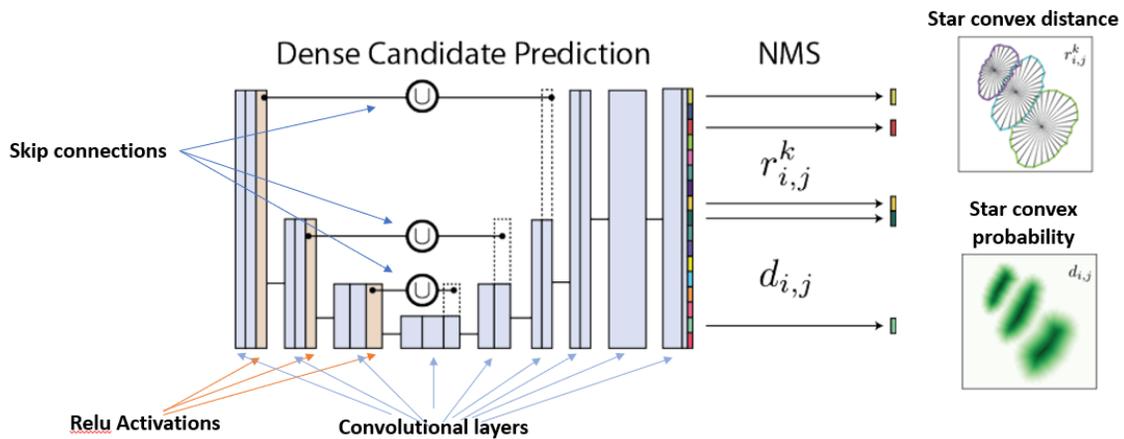

Figure 6: StarDist model architecture [19].

Our initial training model is based on one of the StarDist templates pre-trained on Hematoxylin and Eosin (H&E) nuclei data [31], but it produced poor results in being able to spot the differences between background grid blocks and crystals in the images. To avoid the influence of the bright grid, a preprocessing step was added to the input images in which the brightest 20% of colors in the photo were dimmed by 20%. The model also had difficulty differentiating air bubbles from crystals and detecting individual crystals within clusters. Training with this new data addressed these two problems, and an analysis tool was developed based on the new prediction results. Figure 7 shows the progression from the initial StarDist pre-trained model to the one fine-tuned on the new data.

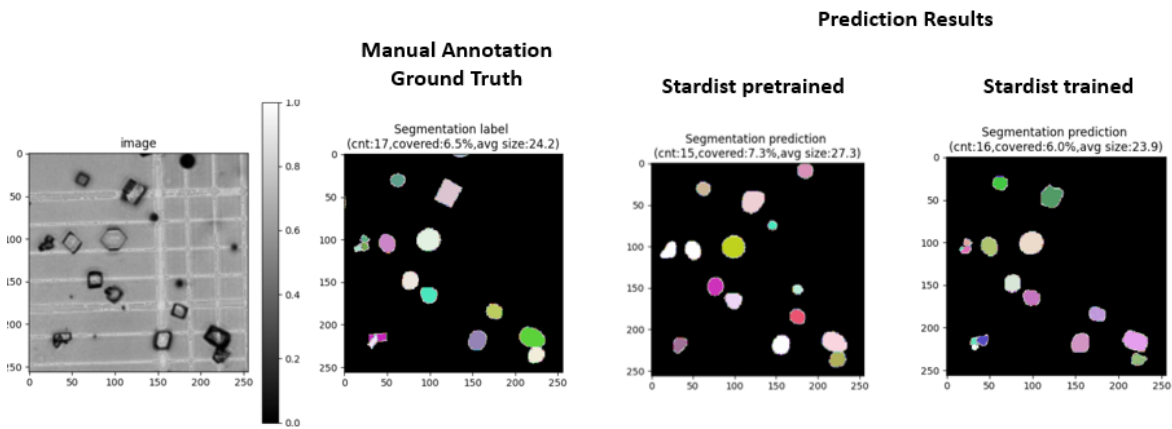

Figure 7: Prediction results example.

*3.3 Developing a usable privacy-preserving imaging tool with MyDataCan*

Having now arrived at a potential solution for the crystal-counting issue, it becomes necessary to guarantee to manufacturers that their proprietary processes (and data) will be protected from competitors.

There are many documented models of this kind of data sharing and, in order to solve for this problem of refusal to share data, a model known as an AI Data Community (AI-DC) has been proposed [20]. This method employs a tool called MyDataCan and, in order to understand how an AI-DC solves for the refusal to share data problem, it's necessary to have a grasp of the tool itself.

MyDataCan ([mydatacan.org](mydatacan.org)) is a privacy-preserving data-sharing platform developed by the Public Interest Tech Lab at Harvard University. Originally designed as a consumer-facing solution to the degradation of online data privacy, the technology was initially predicated on a desire to fundamentally shift governance of personal data away from the companies that monetize it back to the individuals who produce it. In the current data-sharing paradigm, users have minimal knowledge or control over who has access to their personal data, what information they have access to, and how their data is used. Conversely, applications built on top of MyDataCan securely store a copy of users' data in their private "can." Users may choose to edit, privatize, share, restrict, and even delete their data. The platform has been used to secure users' data for contact tracing apps during the COVID-19 pandemic, a nationwide voter registration monitoring service, and a community-building app for college students.

Due to the platform's ability to keep data private and solely in the control of its users, a potential use case for MyDataCan in the manufacturing sector in which manufacturers use the platform to share their data for group analysis via AI or some other research-driven process. Importantly, the platform allows them to do so without revealing their proprietary data or methods to their competitors. In essence, the data *privacy* features of MyDataCan–initially developed to put personal user data in the hands of users–when transferred into the use case of manufacturing, actually become more akin to *confidentiality* features which offer data security while investigating potential solutions to manufacturing issues. In the next section, we will discuss why this is of paramount concern to manufacturers.

### 3.3.1 The importance of data confidentiality for manufacturers

Data confidentiality allows manufacturers to safeguard their innovations, meet compliance requirements, maintain competitive advantage, and build trust with partners, all of which are critical for sustainable growth and resilience in today's marketplace. However, in order to be useful, the dataset must be large enough to train algorithms on and use statistical models to analyze and draw inferences from potential patterns. The resultant challenge is that most small- and medium-sized manufacturers either (a) do not produce the necessary data to train a machine-learning model effectively, nor do they (b) have the workforce capacity to implement and maintain a usable machine-learning model. In our AI-DC model underpinned by the MyDataCan infrastructure (Figure 8), small and medium-sized manufacturers analyze their data using machine learning algorithms and tools built by researchers [20]. Each analysis performed simultaneously helps train and improve the machine learning algorithms, while MyDataCan ensures data privacy and confidentiality throughout the process.

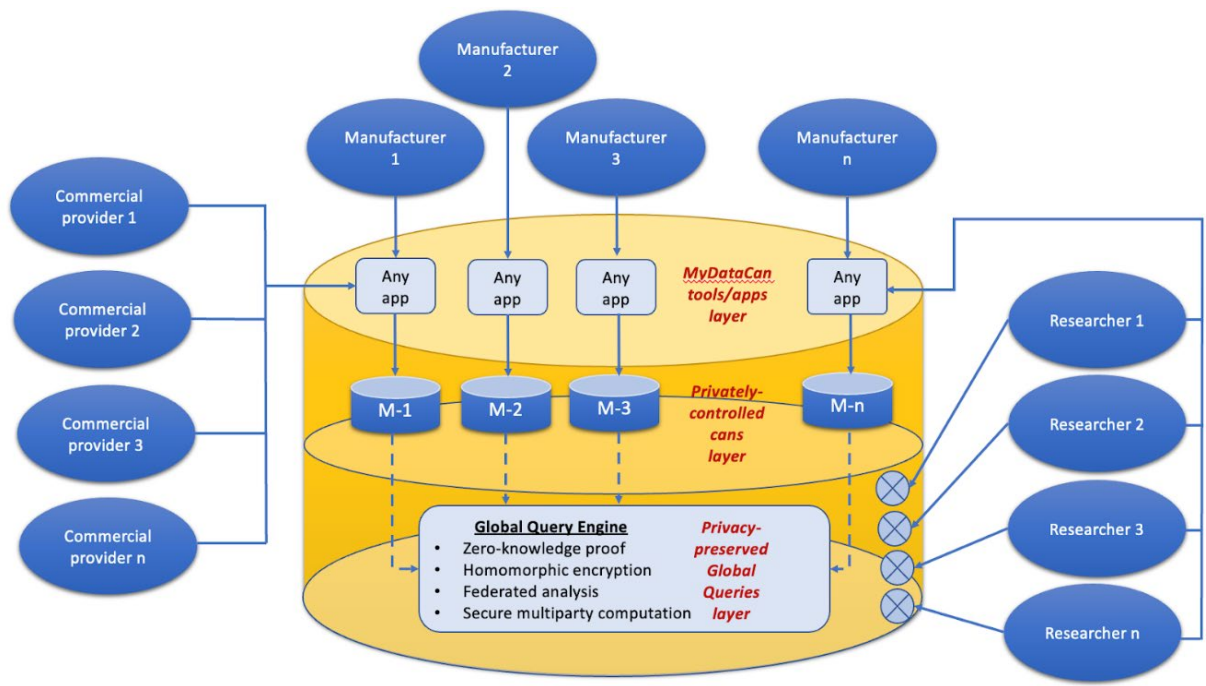

Figure 8: AI Data Community model for manufacturers [20].

In this model, manufacturers upload data to their siloed MyDataCan, analyze it using a machine learning algorithm, and then receive the results back to their MyDataCan. All data is encrypted during the entire process–upload, in transit to/from machine learning tools, during analysis, and while it is being stored by MyDataCan–so that neither the platform itself nor other users can decrypt or access the data at any point in the process. Virtual machines can also be created on demand for broader analysis across multiple data sets. MyDataCan guarantees data privacy, confidentiality, and security using homomorphic encryption, federated learning analysis, secure multi-party computation, and zero-knowledge proofs, as shown in Figure 8. The virtual machine is subsequently destroyed once the analysis is complete to further guarantee data privacy and security [20].

In the following section, we will describe the development of an interface to intake the image data produced by the upgraded crystal counting tool and the construction of a web-based means to connect workers on the factory floor with securely stored data that tool produced.

*3.3.2   Online imaging tool development*

In order to allow any manufacturer to easily use the imaging algorithms in their day-to-day operations, a web-based application was built which could be accessed via an internet browser ([business.mydatacan.org](business.mydatacan.org)). The website was built using Flask–a Python microframework for web applications–which serves the HTML page containing the tool's user interface (which was built using a JavaScript framework called Svelte).

The original crystal counting tool was written as a Python desktop application so required several concessions when translating the code from the desktop interface to the web interface (Figure 9). Initially, high-memory image processing and deep learning algorithms were used to count the crystals without first requiring any pre-processing of the image to remove air bubbles and noise. Hosting those algorithms within the Flask application itself would have put an untenable strain on the website's server, causing the web application to time out and become unusable, so modifications were made to the crystal counting program to run in an Amazon EC2 instance with Python and TensorFlow installed to solve this issue. From there, an AWS Lambda endpoint was created to function as a liaison between the Flask site and the EC2 instance. Instead of displaying the results in a console as the desktop application would, it was modified to return the counting results as a JSON dictionary so we could display them in the user interface (Figure 9).

The user selects the image from their device that they would like to analyze on the user interface and clicks the "Run Analysis" button, as shown in Figure 9. Users may adjust user-defined parameters before running the analysis to help improve the tool's performance. Clicking the "Run Analysis" button triggers an HTTP POST request containing the image data and parameters to a dedicated route in the Flask backend. Flask passes the user's input to the AWS Lambda function and calls the function, which then starts the crystal counting process in the EC2. As the analysis is running, the user interface changes to a "loading" state while the Flask request waits for the Lambda call's result. Once the EC2 finishes the counting process, the Lambda function returns a JSON response to Flask containing the results. Flask sends the JSON results as a response to the user interface's HTTP POST call, and the user interface then displays the results on the page.

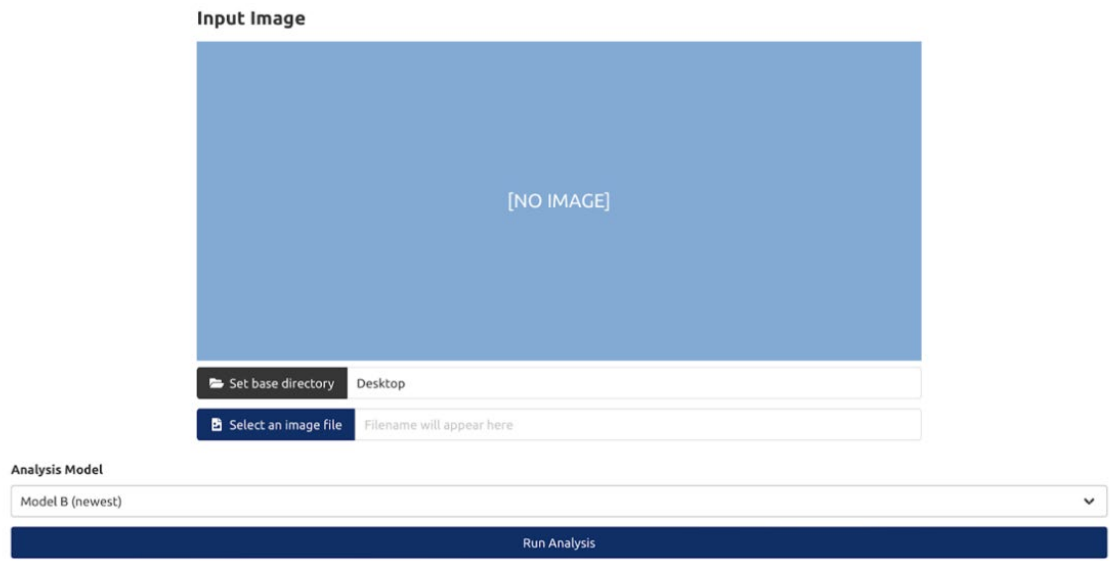

Figure 9: User interface for selecting an image and running an analysis.

Once the ability for users to upload their images to the crystal counter tool was developed, the manufacturer could begin using it in their daily operations. First, the floor manager tested the tool and compared it to manual counting, during which they noted that the image's file name was not persistent after the program had returned the result. This created confusion as to which specific image was being analyzed so, to increase the user's certainty and in accordance with Nielson's UX heuristic of "recognition rather than recall" [21]—the file name was made persistent and visible throughout the entire user experience (Figure 10).

## 4.  RESULTS

In this section, we show the crystal counter tool interface and the image analytic results. There are two versions of interfaces for old camera image analysis and new camera image analysis. The old version interface as shown in Figure 10 includes bubble removal parameters (top two figures), the detailed settings of each parameter are discussed in the baseline model [15]. This interface compares image results with or without bubble removal to verify the parameter settings. In Figure 10, users first click "Use sample data" (as shown in the top left region) in order to upload the images to the web-based app. Once this process is complete, the user then sets the parameters or uses default settings for detection, clicks 'Detect Bubbles,' and then the result is shown on the top right. After the regions of the image containing bubbles are detected, the user can click 'Analyze Crystals' in order to get a crystal size distribution analysis (which appears in the bottom pane of Figure 10).

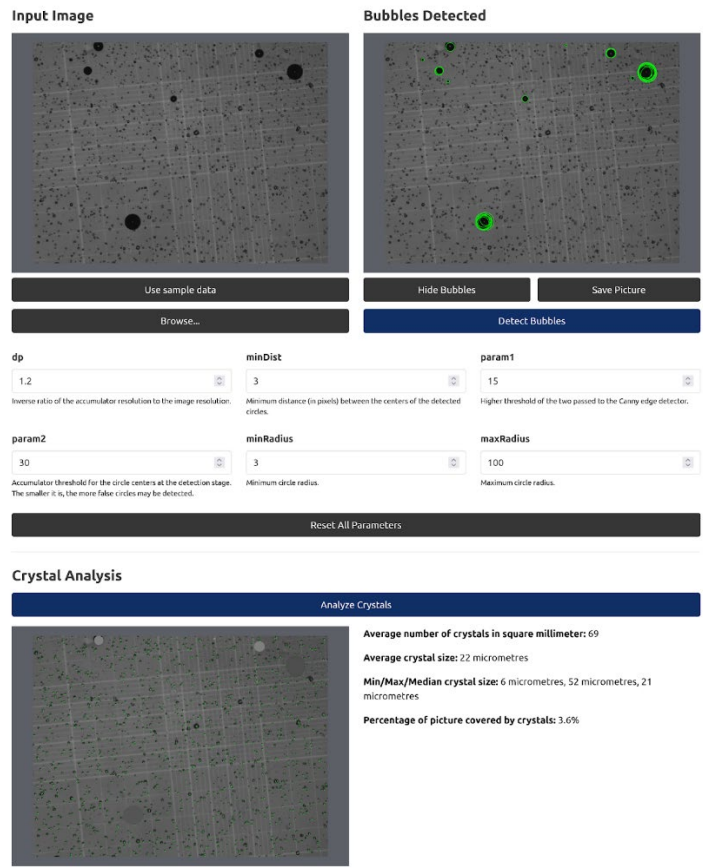

Figure 10: The initial crystal counter tool design. This includes a first step to remove bubbles from the image (detected and highlighted in green at the top of the figure). The operator can then change the image processing algorithm parameters to improve the accuracy of the bubble removal step. Subsequent analysis (bottom of the figure) shows the number of crystals per unit area, percent coverage, and some of their size distribution statistics for the portions of the image in which there are no bubbles.

Figure 11 shows the crystal counter tool interface for the model trained on the new camera data. After this interface was tested by manufacturers, it was decided that there was a need to improve the site's minimal design by changing the user interface's information hierarchy to better emphasize the most important features to their work. According to the operators themselves, the most critical and frequently used information included the image itself and the resulting average number of crystals in the image after running the analysis (referred to as the "seed count"). Based on this feedback, the crystal image was then centered, and the size and prominence of the resulting seed count was increased. Occasionally, the operators would also need to manually count the crystals in the image to verify the algorithm's results, so the analyzed image at its maximum size was placed below the seed count. The crystals were then outlined in a high-contrast color (neon green) compared to the image to make this manual counting process as easy as possible. Lastly, the user-defined parameters were hidden under a toggle at the bottom of the page because it was found that the operators would rarely, if ever, use this feature. Figure 11 shows the crystal counter tool's design after incorporating the operators' feedback.

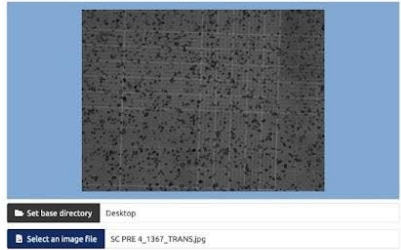
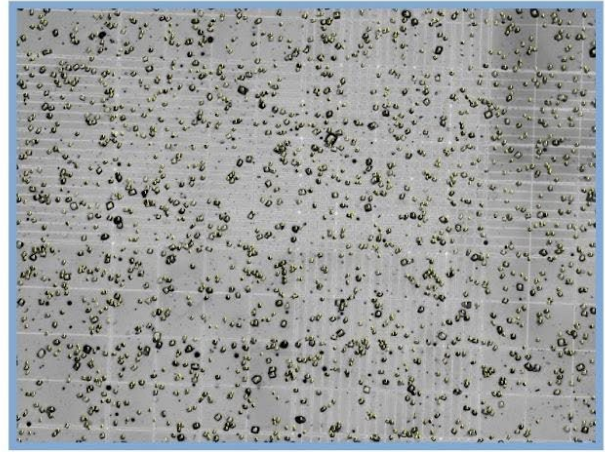
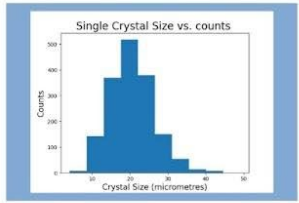
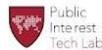

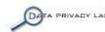

Figure 11: The crystal counter tool design after incorporating the operator's feedback. This tool uses deep learning techniques and doesn't require manual setting of image processing parameters for bubble removal.

## 5. DISCUSSION

This paper proposes and demonstrates a tool for one particular manufacturing process which was successful enough that it has since been adopted by the manufacturer for daily operations, delivering significant benefits such as time savings for operators, enhanced accuracy and consistency in seed counts, and greater confidence in product quality. Before the tool's implementation, operators manually counted seed crystals in just one ninth of each image's total area. This process took approximately five minutes and was performed four times daily by a different person each time. After adopting the crystal counter tool, operators can now analyze the entire crystal image in approximately 30 seconds, equating to a 10x decrease in time spent per analysis. This paper also demonstrates how a privacy-preserving platform, MyDataCan, facilitates the sharing of confidential data with researchers, who in turn build custom solutions which are then deployed on MyDataCan for ongoing real-world use.

*5.1 Future opportunities*

The manufacturer has expressed interest in expanding the usage of the crystal counter tool to other factories, having experienced its value and adaptability. Plans are also underway to apply the tool to later stages of the manufacturing process, where significant crystal agglomeration occurs, but no quantitative or manual analysis is currently performed. In these stages, operators typically provide only qualitative feedback, but the tool enables them to track trends and gradual deviations. The image processing algorithm is also versatile for diverse crystallization applications–such as pharmaceuticals, food processing, or particle processing.

## 6. CONCLUSION

Due to a competitive market, manufacturers are loath to give up their proprietary data. The combination of the flexibility of the process of honing a tool for a specific type of manufacturing process combined with the data-security/privacy-preserving elements of MyDataCan made the undertaking ultimately successful.

The development of a robust crystal counting algorithm to be used on a manufacturing floor represents a significant advancement in both image processing algorithms and data-security/privacy-preservation. The integration of MyDataCan demonstrates that these innovative tools can be deployed in real-world settings, fostering collaboration between researchers and manufacturers, and creating a foundation for future advancements across various industries.

## ACKNOWLEDGMENTS

This work was supported by National Science Foundation Award #2134667, "FMRG: Manufacturing USA: Cyber: Privacy-Preserving Tiny Machine Learning Edge Analytics to Enable AI-Commons for Secure Manufacturing."